\documentclass[10pt,letterpaper,twocolumn]{article}
\usepackage{usenix}
\usepackage{cite}
\usepackage{graphicx}
\usepackage{subfigure}
\usepackage{ulem}

\usepackage[cmex10]{amsmath}
\usepackage{amssymb}
\usepackage{url}

\begin{document}
\title{Experimenting with BitTorrent on a Cluster: A Good or a Bad Idea?}

\author{Liang Wang \quad Jussi Kangasharju\\
  Helsinki Institute for Information Technology\\
  University of Helsinki\\
  Helsinki, Finland\\
  Email: \{Liang.Wang,Jussi.Kangasharju\}@helsinki.fi}

% make the title area
\maketitle

\begin{abstract}
  Evaluation of large-scale network systems and applications is
  usually done in one of three ways: simulations, real deployment on
  Internet, or on an emulated network testbed such as a cluster.
  Simulations can study very large systems but often abstract out many
  practical details, whereas real world tests are often quite small,
  on the order of a few hundred nodes at most, but have very realistic
  conditions. Clusters and other dedicated testbeds offer a middle
  ground between the two: large systems with real application
  code. They also typically allow configuring the testbed to enable
  repeatable experiments. In this paper we explore how to run large
  BitTorrent experiments in a cluster setup. We have chosen BitTorrent
  because the source code is available and it has been a popular
  target for research. Our contribution is two-fold. First, we show
  how to tweak and configure the BitTorrent client to allow for a
  maximum number of clients to be run on a single machine, without
  running into any physical limits of the machine. Second, our results
  show that the behavior of BitTorrent can be very sensitive to the
  configuration and we re-visit some existing BitTorrent research and
  consider the implications of our findings on previously published
  results. As we show in this paper, BitTorrent can change its
  behavior in subtle ways which are sometimes ignored in published
  works.
% Using a real BitTorrent client to verify the research conclusions is
% a popular method in p2p research area. The experiment settings used
% in different papers are also quite different. However, there are few
% serious discussions about the experiment designs and the experiment
% environments. Sometimes, we cannot even find strong arguments about
% why we use such experiment settings. How can we guarantee the
% accuracy of the experiment data.

% Our paper made the initial attempts to discuss these issues in
% details. We show how to do the capacity planning and the importance
% of keeping the experiment design in the "safe region".

% What's more, by running the experiments around the system capacity,
% we observed the how peer selection strategy and piece selection
% strategy affect BitTorrent's behaviors, and what factors can
% restrain their influences. \sout{By careful studying the
% combined-effects from these two mechanisms, we recommend the
% researchers should take the piece selection strategy into account
% when they make analytical models and analyse BitTorrent's behaviors
% in the future research.}
\end{abstract}

\section{Introduction}
\label{sec:introduction}

As networked systems and applications are getting larger and larger in
terms of number of nodes, efficient evaluation methods are needed in
their design. System designers typically have three main methods for
evaluating the performance of their systems: simulations, real world
tests, or emulated tests on a testbed.

Simulations, and analytical modeling where applicable, has the
advantage that very large systems can be evaluated. Modern simulators
can easily reach into system sizes of millions of nodes and analytical
methods can potentially push even further. However, in doing
simulations, the designer is often forced to make several simplifying
assumptions about how the system behaves and thus abstract out many
relevant practical details. Simulators can of course be programmed to
take more details into account, but then their scalability becomes
limited.

Real world tests, for example on systems like
PlanetLab~\cite{planetlab,PlanetLabWeb} offer a realistic network
environment for testing. Although the connectivity and performance of
the nodes in a testbed does not match that of home users, such tests
still have the advantage that the traffic between nodes has to flow
over the real Internet and encounter real network traffic
conditions. The downside of such tests is that they are often limited
in size, with experiments of only hundreds of nodes being
often a practical limit.

Cluster-like testbeds, such as Emulab~\cite{Emulab} or
Grid~5000~\cite{Grid5000}, attempt to strike a middle ground between
simulations and real world tests. They are built as a cluster of
computers, sometimes even geographically distributed, and connected by
a high-speed network. Clusters have the advantage that the actual
system being tested must be written for real, i.e., the same code
would be running on the real network as well. The main issue with
cluster experiments is that the network between the nodes is the
high-speed cluster network, with minimal RTT and practically no packet
loss. However, these can be configured separately, according to
whatever model of a network is being called for. A further advantage
of cluster experiments is that they are typically reproducible, since
the load on the computers and in the network is
controllable~\cite{rao10,antoniu04}. Conditions of real world tests
over the Internet are impossible to reproduce exactly, although
repeating the experiment multiple times will give statistical
confidence. 

% We must mention PlanetLab \cite{planetlab} here, which is a well-known testbed for internet applications in the research area. It is an excellent platform providing us an environment closest to the real internet but still controllable. However, the nodes in PlanetLab are shared by multiple researchers, and different applications will compete for the physical resources. For those very performance-sensitive experiments, it is difficult to guarantee the accuracy of the experiment data. The reproducibility of the experiments is another issue that makes large-scale experiments less attractive on such platforms

% Cluster provides us a closed and exclusive environment for the experiments, so the irrelevant interferences can be minimized to the lowest level. There are some argues that the experiments on the cluster do not take the following things into account, such as heterogeneity in the platforms, bandwidth, RTT, packet loss rate and so on. However, just because of this reasonable simplification in the setting, it is much easier for us to find out the true causes for a problem. Furthermore, such heterogeneous factors can be added to the experiments manually.

Our contribution in this paper revolves around how to design
experiments of network systems and applications on a cluster-like
testbed. As our results show, designing large tests that push the
limits of the cluster is very tricky, with many unexpected effects
cropping up at various places. We cannot stress enough that the
designer must be extremely familiar with both her own system as well
as the underlying operating system and network when running large
tests on a cluster.

As a practical example of a network application we have chosen
BitTorrent in our quest to get the most out of a cluster. We chose
BitTorrent because it is a widely used and studied application and its
source code is available, allowing us to verify some aspects of the
observed behavior from the source code. 

% In our analysis of BitTorrent,
% we observed many situations where existing research has taken
% shortcuts in terms of, e.g., parameter choices, leading to possibly
% unfounded conclusions. A simple reading through the source code would
% have shown the correct parameter choices.

Our starting point is to see how many instances of BitTorrent can run
in parallel on one physical node. We show how to tune and tweak
BitTorrent and present the relationship between the number of clients
per node and other system parameters. Running hundreds of parallel
instances is easily feasible, opening the door to rather large
practical BitTorrent experiments.

We also present analytical means for calculating the overall capacity
(mainly number of clients per node) of an experimental platform. A
commonly used way of looking only at average download rates and times
or aggregated bandwidth turns out to be insufficient. We present a
superior method for calculating the same parameters.

Although experiments on clusters are becoming more common,
e.g.,~\cite{rao10}, there is a lack of general understanding on how
such experiments should be performed and where their limits are. Our
work in this paper provides a first set of answers to these questions
through practical experiments, and therefore provide a recipe for
others to follow when running similar experiments.

This paper is organized as follows. Section~\ref{background} gives
background on BitTorrent. Section~\ref{methodology} describes the
basics of our evaluation methodology. Section~\ref{practical_issue}
shows how we had to tweak BitTorrent and the operating system to get
the most out of them. In Section~\ref{sec:tune-appl-param} we present
the first set of our results. Section~\ref{capacity_planning} presents
the analytical methods for calculating system capacity and
Section~\ref{cluster_nodes} investigates how clients in BitTorrent
cluster under different circumstances. Section~\ref{related_work}
contrasts our results to related work and discusses the implications
of our findings. Finally, Section~\ref{conclusion} concludes the
paper.

\section{Background on BitTorrent}
\label{background}

In this section, we give a brief background refresher on
BitTorrent. It is based on~\cite{cohen03} and the source code of the
Mainline version 4 and version 5 clients.

In order to join a BitTorrent swarm, a peer first needs to obtain the
corresponding meta file, called a torrent file. Then the peer contacts
the tracker whose address is in the torrent file. The
tracker will create a peer list by randomly selecting 40 peers in the
swarm and return it to the requesting peer. With the peer list, the
peer can connect to those already in the swarm and join the
distribution process. By default, a peer will keep connecting to
others until it has 40 connections or \textit{buddies}. After that, it
will stop initiating new connections, but it still accepts connections
from others. When a peer has 80 buddies, it stops accepting new
buddies; any new incoming connections will be dropped immediately. If
the number of buddies drops below a certain threshold, it will
re-request a new peer list from the tracker. So, during the life span
of a peer, it usually maintains 40 to 80 buddies.

The distributed file is cut into \textit{pieces}. The usual size of a
piece can range from 256KB to 1MB,\footnote{This applies to Version
  4. Version 5 determines the piece size in a manner described later.}
but it must be a power of 2. Larger piece size can reduce the size of
a torrent file. When exchanging data, a piece will be further divided
into smaller units, which are called \textit{slices}. In such a way,
the uploads can be pipelined to improve the performance. Slice is the
basic transmission unit.

As one of the core mechanisms, BitTorrent's piece selection strategy
is widely known as \textit{rarest-first}. More precisely, it should be
called \textit{local rarest-first} since the decision is made based on
local information from the peer list. By requesting those rare
pieces, a peer can attract more buddies to download from it. As a
result of tit-for-tat, it will be more likely to be served by others.

Another core mechanism is peer selection strategy.  Leechers (peers
still downloading) and seeds (peers with complete copy of file) have
different peer selection strategy. A leecher will upload to those who
can provide it better download rate, while a seed will upload to
those who can download from it fast. The leecher's strategy is
rate-based tit-for-tat and the purpose is to guarantee the fairness in
the system. Seed's strategy tries to make sure the new replicas will
be generated fast.  Every 30 seconds, a peer selects the buddies to
upload to based on these strategies; others will be choked.

\section{Methodology}
\label{methodology}

In this section, we present the general methodology of our
experiments, including terminology and our experimental environment. 

\subsection{Terms used in this paper}
\label{sec:terms-used-this}

In this paper, we also use the following terms to simplify the
discussion. We refer two connected peers as \textit{buddies}. If a
peer's buddy is on the same physical node with this
peer\footnote{Recall that we run multiple instances per node}, we
refer it as a \textit{native buddy}; otherwise a \textit{foreign
  buddy}. \textit{Aggregated bandwidth} represents the total traffic
generated by a group of peers in every second. It can be further
divided into \textit{aggregated download bandwidth} and
\textit{aggregated upload bandwidth}. In this paper, we only consider
the average value, not the instantaneous one, so the
\textit{aggregated download bandwidth} is calculated as the product of
\textit{average download rate} and the number of peers and likewise
for the \textit{aggregated upload bandwidth}.

All the experiments we performed can be divided into two categories.
In one we set a limit on the leecher's max upload rate; we call this
kind of experiments \textit{upload-constrained experiment}. In the
other kind, we set a limit on the leecher's max download rate, and
call them \textit{download-constrained experiments}. In all of our
experiments, two distinct nodes are used for deploying the tracker the
seed respectively. There is only one original seed in every experiment
and its upload rate is always constrained. Every peer in the swarm
will register itself to the tracker. Our experiment scripts query the
tracker periodically to monitor the number of peers in the swarm.
Only peers that successfully register at the tracker are counted.

% That's the way how we calculate the
% startup peers. So if a peer cannot register itself successfully to the
% tracker, even the BitTorrent instance is running, we won't consider it
% as a successfully startup peer.

\subsection{Methods}
\label{sec:methods}

Our main goal is to understand how far the experiments can be pushed
before hitting the physical limits of the machine. When running
multiple clients on a single physical node, it is vital to know when
the CPU, memory, network, or other factors start restricting the scale
of the experiment. We call the limit below which the experiments still
run without problems the \textit{system capacity}. Any experiment run
above the system capacity limit will yield biased results; thus it is
vital to know that limit when designing experiments.

CPU, memory, or local storage bottlenecks are easy to observe, for
example just by looking at the CPU utilization or memory consumption
statistics. Network bottlenecks are slightly harder to analyze,
especially when multiple peers are running on the same node.

We use the average download rate as an indicator of network
saturation. However, our research shows the average download rate and
the corresponding aggregated bandwidth cannot reflect the system
capacity correctly. The average download rate still remains at a
stable level even though the network has already been saturated. One
key contribution of our work is in analyzing in detail how the
saturation of the network affects the experiments.

% There are several things need to be clarified here. One is the
% system capacity is determined by the minimum capacity of CPU,
% memory, network or any other factors that may restrict the
% experiment scale, in other words, restrict the number of peers
% running on one node. The bottlenecks from the CPU, memory and
% storage are easy to detect. But the bottleneck from the network is
% difficult to handle, if multiple peers run on the same node.

% Average download rate is an important indicator for system
% performance in the study of P2P system. It also plays an important
% role in our experiments.  Intuitively, the average download rate
% should start dropping after the system capacity is reached. However,
% our research shows the average download rate and the corresponding
% aggregated bandwidth cannot reflect the system capacity
% correctly. The average download rate still remains at a stable level
% even though the network has already been saturated. The reason
% originates from BitTorrent's innate characteristic.

We control BitTorrent's network usage by limiting either its upload or
download bandwidth. In most of previous research on BitTorrent, the
researchers only constrain the upload bandwidth and set it to a low
value to model typical home connections, under the assumption that
upload bandwidth is the main constraint in the system. This kind of a
setting has a serious weakness, because the standard BitTorrent
version 4 client has no enforced download rate limitation; it is
limited by the available physical bandwidth. In a heterogeneous
network, controlling only the upload bandwidths leaves open the
possibility of clients downloading at rates exceeding their physical
bandwidth.

In our experiments we try different values for upload bandwidth, but
most of our experiments are run with a high value of 40~Mbps (5
MB/s). Although such upload bandwidths are still rare on the Internet,
using a high value allows for easier probing of the system capacity
limits. Using a high upload bandwidth, we can guarantee that the
network will become the first bottleneck. This has the added benefit
of allowing us to observe BitTorrent's reactions to changes in network
conditions more easily. As a result, we are able to examine how
BitTorrent's piece and peer selection algorithms interact with each
other and get more insight on how peers cluster in a BitTorrent swarm.

\subsection{Experiment Environment}

Our experiments are performed using nodes equipped with a 8-core
2.8GHz CPU, 32GB memory and connected to a Gigabit Ethernet. The
underlying operating system is Ubuntu SMP with Linux 2.6 kernel. The
TCP congestion control used in the network between the nodes is CUBIC
TCP. The parameters \texttt{net.ipv4.tcp\_wmem} (controls the sending
buffer) and \texttt{net.ipv4.tcp\_rmem} (controls the receive buffer)
are set to "4096, 16384, 4194304" and "4096, 87380, 4194304"
respectively (minimum, default, and maximum). We observed a slight
performance increase if the default sending buffer was increased to
64~KB, but kept it at 16~KB for our experiments.

% \textbf{JUSSI: EXPLAIN THE
%   MEANING OF THESE SETTINGS}

%   \textbf{LIANG: For each tcp connection, wmem controls the sending
%   buffer, which is actually the size of cwnd; while rmem controls
%   the receiving buffer, which is the size of rwnd. These two
%   parameters are very important for tuning tcp to gain better
%   performance, especially when the bandwidth delay product is large,
%   such as in high performance cluster environment. The usual
%   strategy is increase the sending and receiving buffer
%   reasonably. Tuning these parameters is the administrator's task,
%   :) The three values represent 'minimum', 'default' and 'maximum'
%   respectively. For details, please refer
%   \url{http://wwwx.cs.unc.edu/~sparkst/howto/network_tuning.php} }

%   \textbf{In HIIT's cluster, the default sending buffer is
%   16KB. Actually, it would be better to increase it to 64KB. I
%   remembered there would be some improvement in the performance, but
%   not significant improvement.}

The BitTorrent client we used is the BitTorrent Mainline Version 4
client, with some local modifications as detailed below. The code for
our modifications and experiment setup are available at
\texttt{http://www.cs.helsinki.fi/u/lxwang/p2p}

% For the BitTorrrent client used in our experiment, we considered
% several ready-made clients. However, they are not full-fledged, and
% cannot satisfy our requirements. For example, the max download rate
% cannot be set, data logged is not comprehensive enough. What's more,
% all of them are only for the small-scale experiments, are not
% qualified for the use in large-scale and heterogeneous experiments.

% So, we modified client by ourself, the target client is official version - BitTorrent Mainline Ver.4

\section{Tweaking and Tuning}
\label{practical_issue}

In this section we present our modifications to the BitTorrent client
and discuss how the operating system had to be tuned to allow for the
largest number of clients running in parallel. We will present the
details of BitTorrent parameter settings in
Section~\ref{sec:tune-appl-param}. 

\subsection{Running Multiple Peers on One Node}
\label{sec:runn-mult-peers}

% \sout{[Give a table here, show different combinations here, products are the same, avg dl as benchmark; show running multiple peers per node is reasonable]}

The original design of BitTorrent only allows one instance running on
one node. We considered the possibility of using virtual machines but
decided against them because of their relatively high resource usage
which hundreds of parallel VMs would engender.

% We also considered virtual machines like VServer and KVM. One advantage of virtual machine is they can provide strong isolation at very low level. Some hardware level parameters such as IP address, physical upload and download bandwidth can be configured for each peer respectively. This is a very charming feature when deploying heterogeneous experiments. However, virtual machines are relatively resources-consuming. The system performance degrades fast if we use this way. Virtual machine can be a good choice only if very few peers are deployed on one machine such that the experimenter can guarantee the system performance will not degrade.

% Considering the configuration of the nodes in our cluster, one peer
% per node is really a waste of our resources! What's more, even we
% have hundreds of machines, they are still not enough if we want to
% enlarge the scale to thousands and even ten thousands of peers. So
% running multiple peers on one physical node is inevitable. With the
% capability of running multiple instances on one node, it is possible
% to deploy large-scale experiments with limited nodes. If there are
% large amount of nodes are available, as we are going to show in the
% following sections, the experiment scale can be easily enlarged
% without even considering the loopback restrictions, and the capacity
% planning is easier to handle.

The simpler solution was to modify the BitTorrent client to allow
multiple instances run in parallel on a single machine. The resident
memory for each instance is 10---14MB, so the memory will not be a
bottleneck. We also added some functions such as creating working and
configuration directories on the fly to avoid conflicts between the
instances. BitTorrent also has a built-in limitation to allow only one
connection per IP address, which we disabled. (The limitation is
intended to prevent free-riding clients from creating several peer IDs
and pretending to be multiple peers; this is not an issue for us.)

% As our future study mainly focuses on higher level strategies such
% as peer selection or piece selection. Isolation on the application
% layer is quite enough for us. In order to support multiple
% instances, we modified the code and added some functions such as
% creating working and config directories on the fly to avoid
% conflicts. The resident memory for each instance is 10$\sim$14MB, so
% the memory won't be our bottleneck.

% We must point out one side-effect caused by running multiple peers
% on one node, even though it is not a problem in our research. In
% BitTorrent, a peer uses a 20-byte long peer-id to identify itself
% uniquely. A greedy peer may try to make multiple connections to
% another peer by using different peer-ids. In such a way, it can get
% better download performance. To prevent such things, BitTorrent
% allows only one connection from one ip by default.

% In order to run multiple instances on one machine, if all of them
% share the same ip, this feature has to be disabled. Since all the
% clients running on the cluster are under our control, this feature
% is not useful to us. But if it is important in your experiment, it
% is better to know how it affects BitTorrent's behaviours.

\subsection{The Logger Module}
\label{sec:logger-module}

We implemented a Logger module in the client. The Logger module is
used to collect important information during the lifespan of a peer in
the system. It will record the important events happening within the
client, such as the timestamps for starting the client, joining the
swarm, finishing downloads, leaving the system and so on. Besides
that, the Logger module also takes a snapshot of the system every
second. The snapshot includes information such as, the current upload
and download rate, share ratio, transferred data size, and the
connections maintained by the client at the moment.

Since the Logger module records almost all the important information,
it gives us a good way to study the BitTorrent behavior in
detail. Of particular benefit is the ability to track connections,
which is needed when investigating peer selection strategies.

%  Especially the ability of connection tracking, which is very
% helpful in study of peer selection strategies. 

% Our instrumented
% BitTorrent also supports some other useful features. For example, it
% is able to mimic the peers behind a firewall or the free-rider's
% behaviours with various switches.

\subsection{Bypass I/O Operations to Hard Disk}
\label{sec:bypass-io-operations}

Our first experiment was performed in a simple setting:
one seed and one leecher. Since we did not limit the upload or
download rates, the transfer rate should reach somewhere close to the
network bandwidth of 1~Gbps or 125~MB/s. However, the stable transfer
rate in our experiment was only 70MB/s, far below the value predicted.

The bottleneck turned out to be the I/O operations. Writing the
received file to hard disk cannot keep up with the speed at which the
client is receiving data from the network and lot of CPU resources are
wasted in I/O wait. 

% By monitoring the system carefully, we found the bottleneck is from
% I/O operations. The writing speed to the hard disk cannot keep up
% with the speed at which BitTorrent receives data. Lots of CPU
% resources are wasted on I/O wait. \sout{ When running multiple peers
% on one node, there will be higher aggregated bandwidth, and write
% operations to the disk must introduce more overheads. } So,
% preventing BitTorrent from writing to the disk is crucial to gain
% better performance. We also have another good reason for doing so,
% which comes from the limitation of storage capacity. In order to get
% more accurate data, we use large file (5GB) as distribution
% content. However, the hard disk cannot provide us enough space if
% every peer really writes data to the disk.

We considered two ways to bypass disk writes:
\begin{enumerate}
\item Simply throwing all received data away would eliminate all
  writes, but the client must be able to serve other peers with the
  correct data, so the file has to be available to it.
\item Storing the file in memory would help with I/O, but we do not
  have enough memory for hundreds of peers keeping a
  file of several GB in memory.
\end{enumerate}

Our solution is a combination of the two methods. We intercept read
and write operations within BitTorrent. When writing, we simply
discard all data and when reading, we configured the client to read
from a single, shared file. Using a shared file also means that the OS
is likely caching the file in its buffers, since all clients regularly
access all parts of it, but we only have one cached copy as opposed to
each client having its own copy. We pre-load the file before the
experiment, to allow the OS to cache it without affecting the
beginning of the experiment.

As a result, we are able to eliminate I/O wait almost
completely. Table~\ref{tab:table_diskio} shows the performance of the
simple scenario with and without I/O bypassing. It also turns out that
even when we run hundreds of clients per machine, the CPU resources
spent in I/O wait are close to zero.

% We let BitTorrent do nothing
% in write operation requests, just drops the received data. At the same
% time, all the read operations from different instances are redirected
% to the same file. Since there is only one copy of the distribution
% file, the storage space is saved and issues above are solved.

% Considering all the nodes are equipped with the large memory, we let the operating system cached the complete file in a experiment. We also adopt memory-mapped file mechanism in the instrumented client, and make the cached file shared by all the peers on the node. In order to eliminate the major page-faults because of the first access to the file, the file is pre-loaded before the experiment. With these methods, overheads caused by page-faults and system-calls can be reduced. The CPU resources spent on I/O wait is almost zero even hundreds of peers run on the same node.

% Then we repeated the experiment with the simplest setting above, the transfer rate increased from 70MB/s to more than 110MB/s. The improvement can be seen in the table \ref{tab:table_diskio}.

\begin{table}[!tb]
\centering
\begin{tabular}{|p{1.8cm}|p{2.5cm}|p{2.5cm}|}
\hline
I/O bypass &  Transmission rate & CPU on I/O wait \\
\hline
NO & 75MB/s & 85\% \\
\hline
YES & 115MB/s & almost 0\% \\
\hline
\end{tabular}
\caption{Average download rate with and without I/O to hard disk}
\label{tab:table_diskio}
\end{table}

\subsection{Restrictions from the OS}

Our next experiment was to test the maximum peers we can start on a
node. The goal was to identify possible limits in OS or BitTorrent on
starting multiple clients. Since our goal was to find out the system
capacity limits, we simply started all clients at the same time.

% considering the degradation of system performance. The
% purpose is to figure out what other factors might restrict the maximum
% peers on a node.

One restriction is from BitTorrent itself. By default, BitTorrent
tries to listen on port 6881 for incoming connections. If port 6881 is
occupied, it will try the others in the range 6881--6999
sequentially. This means we can only start 119 peers, after which
BitTorrent will report an error. So we simply extended this range to
6881--9999 to guarantee enough ports.

We also observed an unexplained limit on the number of BitTorrent
clients we were able to start (quasi) simultaneously. After starting
700 clients, the speed of starting new processes slowed down and after
800 clients it practically stopped. We were not able to get more than
835 clients started in this manner. We investigated several
possibilities, but were not able to find a cause for this behavior. It
was not an OS limit on starting processes, filehandles, available
local ports, nor the tracker. The behavior is repeatable, but so far
we have not been able to find the cause.

In practical terms, this means that we have a hard limit on the number
of peers that can start ``simultaneously''. If an experiment tries to
capture realistic arrival patterns, this is not necessarily an issue,
but it is something the designer should keep in mind. In our
experiment setting, we were able to start 500 clients on a single node
within 15 seconds.

% Start up speed is also very crucial to the experiments. Since the
% peers' arrival pattern in future experiments will be based on the
% simulation of the arrival pattern in the real world. Our experiment
% scripts can start the peers at different time in the experiment
% according to the configuration file. If certain amount of peers are
% supposed to join the swarm at the same time, there should not be too
% much time difference between the first started peer and the last. In
% our experiment settings, 500 peers can be started within 15 seconds
% on one node.

Besides the above restrictions, there are also some others from the
kernel and TCP, such as the maximum processes a user can start,
maximum sockets, queue length for loopback interface,
\texttt{tcp\_max\_syn\_backlog} and so on. All these parameters have influences
on the experiments and system performance. An experimenter should be
very careful when he decides running multiple peers on one node,
especially when the experiments are performed near the system
capacity. Most probably, the experiment may be overwhelmed by tons of
underlying details and parameters. However, knowing these restrictions
enables us to control the experiment completely. We did experiment
with tuning the kernel and TCP and did observe small potential
performance gains, but none were significant enough to merit the added
trouble of tweaking them.

% \textbf{JUSSI: WHAT
%   DID WE DO ABOUT THESE? HOW RELEVANT WOULD THEY BE FOR US?}
  
%   \textbf{LIANG: Some systems have a limit on the maximum sockets a
%   user can create, or maximum processes a user can start, these can
%   restrict how many instances can be started up on a
%   node. tcp\_max\_syn\_backlog is the maximum queue length for
%   half-open connections, it can be used to prevent SYN flood
%   attack. It can also be tuned to improve tcp performance on
%   Linux. And I also tried increasing the queue length for loopback
%   device, in other words, increase the buffer size for loopback
%   device. The average dl rate can also be improved a little bit, but
%   not significantly. What's more, the queue length for loopback is
%   NOT the bigger, the better. All these parameters are quite general
%   to the TCP or system performance, not specific to BitTorrent. They
%   do have some effects on average download rate, but the effects are
%   marginal, if I may say. The point is if an experimenter tries to
%   run too many peers on a node, and tries to control all the
%   relevant parameters, even those have marginal effects, then it
%   would be an endless nightmare.}

\subsection{Other Issues}
\label{sec:other-issues}
%\subsubsection{Impacts on the performance from piece size}

When running multiple instances on one node, the piece size also has
an impact on the performance. In Mainline Version 4, BitTorrent uses a
dictionary to manage all the pieces. The smaller the pieces are, the
more items will be in the dictionary, and more overhead will be
introduced. Version 5, on the other hand, decides the piece size as a
function of the file size and does not allow for more than $2^{12}$
pieces. We created the torrent files on Version 5 in order to get more
clients per node because the torrent files of Version 5 are
smaller. However, we decided to use Version 4 in our experiments
because it is written in a clearer way and much easier to adapt and it
is able to use the torrent files created by Version 5. Furthermore,
the basic structure and core mechanisms are the same in both
versions. We also experienced problems with unexplained, incomplete
downloads with Version 5.

We observed also another interesting property of how Version 4 manages
the downloads. It keeps track of the pieces in a hash table. When we
tried to use an all-zero file as the file to be distributed, we
observed a hit in performance, because all the pieces have the same
hash and the dictionary manager had to resolve all those hash
collisions. Hence, the important lesson to learn from this is to use a
''normal'' file in experiments. We have not seen clear evidence of
previous research falling for this, however it is good to note it.

% In a test version of our deployment script, we let the script
% automatically generate an all-zero file as distribution
% file. However, we found the system reached its limit much earlier
% than before. By monitoring the CPU usage, we found using all-zero
% file will generate more CPU load. The reason is all-zero file will
% result in many pieces with the same SHA-1 code. What's more,
% BitTorrent uses dictionary manage these pieces while downloading,
% many duplicated keys cause the confusion.

% Actually, the pieces with the same key should be organized as a list
% and saved into the dictionary under that key. However, BitTorrent
% doesn't do in this way, at least I cannot see this in the official
% version 4. Considering Mainline version 4 is widely used in research
% area, it would be better not to use all-zero files, or any other
% files with too many duplicated pieces as distribution content.

\section{Setting BitTorrent Parameters}
\label{sec:tune-appl-param}

BitTorrent has several dozens of parameters that can be tuned, some of
which have great influence on the performance. Many developers spend
quite a lot of time on tuning and testing those parameters to gain
better performance, and these parameters are set to different values
in different implementations. Even in the official implementation,
same parameters are changed in different versions.  These changes on
the parameters reflect the changes in the network environment, at
least from the implementer's perspective.

Basically, BitTorrent is designed for low bandwidths and some
parameters which give BitTorrent good performance on the Internet are
not suitable in a high performance cluster. Hence, we need to tune
these parameters carefully to obtain the maximum number of clients per
node. A cluster is somewhat of an artificial environment and we need
to be careful when generalizing the results to other scenarios. We
believe the applicability of the results obtained on a cluster depends
on what are the metrics of interest and how they behave. For measuring
the effects of ``high level'' behavior (e.g., piece selection, etc.),
we believe the results obtained on a cluster can be considered
representative. On the other hand, applications with strict timing
requirements, e.g., streaming, would not get representative results on
a cluster. The exact extent to which experiment results from a cluster
can be generalized is part of our future work.

% What's more, the impacts
% from these parameters can be amplified when we run multiple peers on
% one node. By carefully tuning them, we can deploy more peers on a
% node. We investigated the impacts from various parameters, and below,
% we will talk about 3 important ones.

All in all, we investigated many different parameters and below we
explain the effects of the 3 main parameters we discovered.

\subsection{Sending buffer}
\label{sec:sending-buffer}

The first parameter that can be tuned to improve the performance is
\texttt{upload\_unit\_size}. It controls the sending buffer in the application
layer. When BitTorrent sends data, it writes 1380 bytes into TCP layer
every time by default, and as a result, it generates a large amount of
I/O operations in our experiment setting(high transfer rate, multiple
peers on one node, etc.). However, when BitTorrent receives data, it
will read up to 100KB from the TCP buffer every time.

By increasing this \texttt{upload\_unit\_size}, more data can be passed to TCP
layer in a single write operation and the number of I/O operations can
be reduced for a given amount of data. In our experiments, we
increased this number to 64~KB, which we observed to give large improvements.

\subsection{Slice size}
\label{sec:slice-size}

As mentioned in Section~\ref{background}, slice is the basic
transmission unit. If a slice is corrupted, the whole slice needs to
be re-transmitted and thus a large slice size can be inefficient. In
the official version the parameter \texttt{download\_slice\_size}
controls this and is 16~KB in Version 4 and 32~KB in Version
5.\footnote{Version 5 actually calls it
  \texttt{download\_chunk\_size}, but its effect is the same.} The
reason for the increase in slice size between the versions is due to
the increased bandwidths on the Internet, since larger slices are more
efficient and faster network connections do not penalize the use of
larger slices.

We experimented with several slice sizes and found out that increasing
from 16~KB to 32~KB yields a significant improvement and a further
increase to 64~KB resulted in a clear improvement over 32~KB. Further
increases beyond 64~KB did not yield much improvements, so we decided
to use the value 64~KB for the slice size.

% \textbf{JUSSI: WHAT VALUE DID WE USE?} 
% \textbf{LIANG: 64KB, you can check nl\_10.png in graph folder, it is the result from previous experiment. I fixed at two points (75 peers/node and 85 peers/node) where the system is already overloaded, then increased the slice size. You can observe there is significant jump when slice size increased from 16KB to 32KB. According to results from many experiments, 64KB can give better performance than 32KB, even it can not be observed from nl\_10.png. Then, keep increasing the download slice won't do much improvements to the avg dl rate any more.}

\subsection{Concurrent uploads}
\label{concurrent_uploads}

The number of concurrent uploads\footnote{also referred as upload
  slots} plays an important role in BitTorrent's clustering behavior.
The larger this value, the more difficult it is to see clustering of
peers. In the extreme, when a peer uploads to all of its buddies, it
is practically impossible to see any clustering. In many BitTorrent
implementations, a user can set this value explicitly, but if this
value is not specified, BitTorrent will calculate it based on the
maximum upload rate, as the equation \eqref{eq:uploads} shows.
$uploads$ denotes the number of concurrent uploads, $rate$ denotes the
maximum upload rate in KB/s. When $rate$ is set to negative, it means
no limits on the upload rate. We can see that when the max upload rate
is unlimited, 7 upload slots will be used, which is quite a
conservative number.
%  This
% can be viewed as an implication that BitTorrent is designed for
% low-speed network.

\begin{equation}
  \label{eq:uploads}
  uploads = \left\{
    \begin{array}{rl}
      2 & \text{if } 0< rate < 9,\\
      3 & \text{if } 9 \leqslant rate < 15,\\
      4 & \text{if } 15 \leqslant rate < 42, \\
      \sqrt{rate \times 0.6} & \text{if } rate \geqslant 42, \\
      7 & \text{if } rate \leqslant 0.
    \end{array} \right.
\end{equation}

Concurrent uploads also has strong influence on the system capacity.
The larger the number of concurrent uploads, the smaller the overall
system capacity, because too many concurrent uploads also cause a
large amount of I/O operations.

\subsection{Peer Set Cardinality}
\label{sec:peer-set-cardinality}

In many published works about BitTorrent, researchers do not care much
about how many peers they use. However, our results show that
selecting a too small number of peers can have a strong effect on the
results. 

% "What is the minimum peers we should use in an experiment?" - is an issue worth discussion. It seems the researchers don't show much concern on the minimum peers we should use in an experiment, while they try their best to enlarge the experiment scale. However, our research shows the minimum peer set cardinality can also affect the experiment data.

We found that as the swarm size grows from 0 upwards, the average
download rate keeps on decreasing until there are 40 peers in the
swarm. Then the average download rate will remain roughly constant
until we hit the system capacity limit. The reason for this behavior
is that the peer list that a peer gets from the tracker contains 40
peers. Hence, when the swarm has less than 40 peers in total, every
peer knows every other peer and the connection graph between them is a
full mesh. This means that every peer has to maintain more buddies and
thus the overhead of maintaining the connections increases. In large
swarms, peers only maintain connections to about 40 peers, so the
overhead remains stable after that point, until we reach the system
capacity. 

% According to the data from our experiments, we found that the
% average download rate keeps decreasing until there are as many as 40
% peers in the swarm. Then the average download rate will remain
% stable even more peers join in (within the system capacity). The
% reason is the minimum peer set maintained by a peer is decided by
% the peer-list it received from the tracker. By default, the tracker
% will return a peer-list containing 40 peers. So, if there are no
% more than 40 peers in the swarm, they can construct a complete graph
% since all of them will appear on each other's list. More peers
% joining in means a peer has to maintain more buddies and more
% overheads are introduced because of the communication among
% them. However, after there are more than 40 peers in a swarm, more
% peers joining in will not increase the burden of a peer, since the
% peer-list can contain only 40 peers as before. The average download
% rate will enter into stable stage if the system limit is not
% reached.

The important lesson to learn from this is to make the swarm size in
any experiment larger than the peer list.

% Many papers design the experiments without caring this minimum
% number, and some use much less peers than the minimum peer set
% cardinality. We recommend the minimum peers should not be less than
% that a peer-list can contain. Or the experiment data will not be
% considered ad representative.

\subsection{Effect of Tuning}
\label{improved_result}

\begin{figure}[!tb] 
  \centering 
  \subfigure[Average download rate]{\label{tu_1a:a} 
    \includegraphics[width=7.5cm]{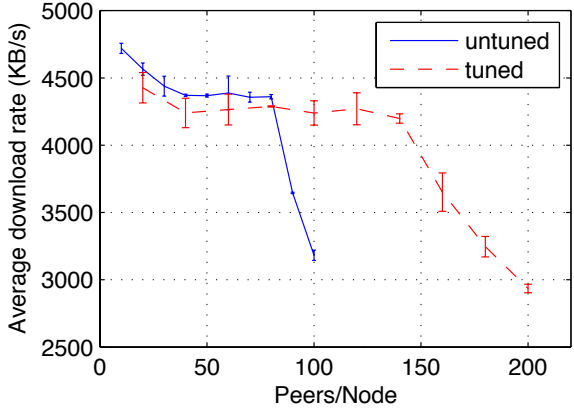}} 
  \subfigure[Aggregated download bandwidth]{\label{tu_1a:b} 
    \includegraphics[width=7.5cm]{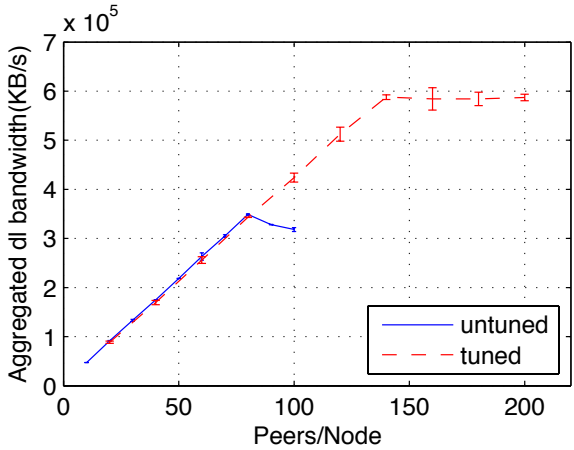}} 
  \caption{Effects of tuning BitTorrent parameters on average download
    rate and aggregated bandwidth as a function of peers per node in
    download-constrained experiments. The bars show 99\% confidence
    intervals.}
  \label{fig:effect-tuning}
\end{figure}

Figure~\ref{fig:effect-tuning} shows how many peers we can deploy on a
single node with and without tuning the parameters as described
above. In these experiments we had one seed on a different machine
with a maximum upload rate of 5~MB/s and all the leechers were on
another machine and we constrained the leechers' download rates to
5~MB/s. Upload rate was unconstrained and the number of concurrent
uploads for leechers was 7. File size was 2~GB. Figure~\ref{tu_1a:a}
shows the average download rate and Figure~\ref{tu_1a:b} shows the
aggregate download bandwidth in the system.

We can see that the average download rate in the \textit{untuned case}
enters into the stable phase at 40 peers/node (recall that the peer
list has 40 peers, as discussed above), and remains stable till it
reaches 80 peers/node. After 80 peers/node, the average download rate
drops sharply. This change can also be observed clearly on the
corresponding aggregated bandwidth in Figure~\ref{tu_1a:b}. Before
reaching the system capacity at 80 peers/node, the aggregated
bandwidth keeps increasing linearly to 350MB/s. Without tuning the
parameters, we can therefore deploy a maximum of 80 peers on a node.

Looking at the curves for the \textit{tuned case}, we see that the
average download rate remains stable until we have 140 peers per node,
almost double of that of the untuned case. Likewise, the aggregated
bandwidth can reach almost 600~MB/s. In other words, by properly
tuning the parameters we can deploy around 140 peers on a node at
maximum. The tuned case exhibits similar behavior to the
download-constrained case, with a limit of about 140 peers per node.

\begin{figure}[!tb] 
  \centering 
  \subfigure[Average download rate]{\label{tu_3a:a} 
    \includegraphics[width=7.5cm]{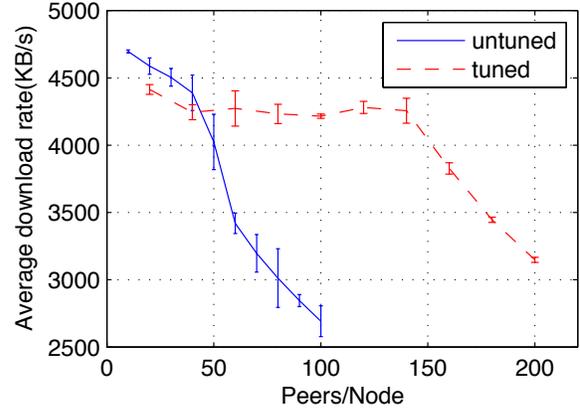}} 
  \subfigure[Aggregated download bandwidth]{\label{tu_3a:b} 
    \includegraphics[width=7.5cm]{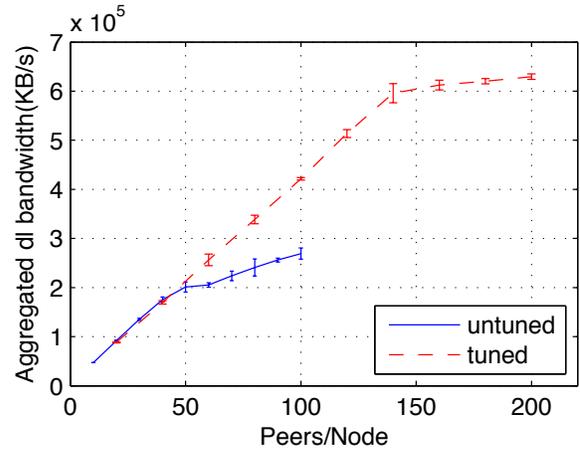}} 
  \caption{Effects of tuning BitTorrent parameters on average download
    rate and aggregated bandwidth as a function of peers per node in
    upload-constrained experiments.}
  \label{fig:effect-tuning-upload} 
\end{figure}

Figure~\ref{fig:effect-tuning-upload} shows the results for
upload-constrained experiments, where every leecher had unlimited
download rate, but the upload rates were limited to 5~MB/s; seed's
upload was also limited to 5~MB/s as in the previous experiment. The
results are similar to the ones from the download-constrained
experiment, with the difference that the average download rate for the
\textit{untuned case} never enters a stable phase; instead, after the
initial decrease in swarms under 40 peers, it continues to decline,
indicating that the system is already overloaded. Interestingly, the
aggregated bandwidth still keeps increasing slightly after 40 peers,
but as the average download rate indicates, the system is overloaded
and the experiment is no longer correct.

Note that the tuned case uses a fixed number of 7 upload slots in both
cases. In the upload-constrained case we should actually let
BitTorrent decide the number of upload slots according to
equation~\eqref{eq:uploads}. We tried this as well (results shown
later in Figures~\ref{cap_1:c} and \ref{cap_2}) and observed that the
behavior is practically identical up to the capacity limit. After the
capacity limit has been reached, 7 upload slots means a more stable
system performance with a very small variation in average download
rate and aggregated bandwidth whereas the real value of 54 slots
results in highly variable behavior (Figures~\ref{cap_1:c} and
\ref{cap_2})

% \textbf{LIANG: Sorry, again, I failed to make it very clear. In tuned case, the concurrent uploads are explicitly set to 7; but in untuned case, I let BT calculate the concurrent upload by itself according to \eqref{eq:uploads}. That is why in tuned case, the curve shapes are almost identical in both upload- and download-constrained experiments. However, in untuned case, I didn't set concurrent uploads explicitly, but let BT calculate it by itself. So, the curve shape became quite different. Since in download-constrained experiments, the concurrent uploads is 7, while in upload-constrained experiments, the concurrent uploads is 54. That explains why avg dl rate never enters into stable stage in upload-constrained experiment in untuned case. Because too many concurrent uploads generate large amount of I/O, recall \ref{concurrent_uploads} }

Figures~\ref{tu_1a:a} and \ref{tu_3a:a} show that the average download
rate for the tuned case is slightly lower than untuned case when the
number of peers is very small. This is because the tuned case uses a
larger slice size, hence a piece will be divided into a smaller number
of slices. Request pipelining which allows efficient parallel
downloads is not as efficient in this case, hence the average download
rate suffers slightly, but we have less I/O overhead.

\section{Capacity Planning}
\label{capacity_planning}

After determining how the parameters are to be tuned for the best
performance, we now turn to the more general issues related to
capacity planning. Our goal is to determine general rules of thumb
which a system designer can use to evaluate the performance of the
system. 

% After obtaining satisfactory performance in the previous experiment,
% we started our investigation in capacity planning. Capacity planning
% is very crucial for large-scale experiments on the cluster. The
% purpose is to ensure the experiments are performed within the system
% capacity. However, figuring out the accurate system capacity is not
% an easy job. In this section, we will show how we estimate the
% system capacity to guide the design of the experiments.

First we show that the naive method of only looking at the average
download rate is not sufficient, and then we turn to a more elaborate
mechanism for estimating the system capacity limit.

% We start from a very simple experiment setting in the first subsection. We will show how to decide the system capacity in a naive way. In the second subsection, we will show a better way to calculate the system capacity and claim the method used in the first section is immature. The actual experiment scale is much smaller than the value estimated with naive method. The average download rate cannot be used as the only criterion.

\subsection{Naive Method For Capacity Planning}
\label{sec:naive-meth-capac}

In the naive method, we only take average download rate into
account. If there is no significant drop in average download rate,
then the experiment is considered to be reasonable.

First we experimented with placing all the leechers on a single
node. We increased the number of leechers until the average download
rate was no longer stable. All leechers were upload-constrained and we
used different values for upload bandwidths: 10, 20, 40, and
100~Mbps.

% Experiment settings: only one node was used for deploying
% leechers. The number of leechers was increased step by step until the
% average download rate was not stable any more. We ran several rounds
% of experiments with different max upload rates: 10Mbps, 20Mbps, 40Mbps
% and 100Mbps.

First, we investigated whether we can use the simple formula
$y=\frac{a}{x}$ to roughly estimate how many peers we can put on a
single node. $y$ is the maximum number of peers we can put on a single
node, $x$ is the maximum upload or download rate we set, and $a$ is a
constant related to the aggregated bandwidth. If the transmission rate
and maximum number of peers on a node exhibit this simple relation,
then we need not redo the capacity probing every time we change the
experiment settings.

\begin{figure}[!tb] 
\centering 
\subfigure[Upload bandwidth 10 Mbps] { \label{cap_1:a} 
\includegraphics[width=5cm]{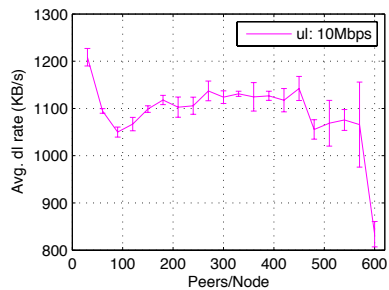} 
} 
\subfigure[Upload bandwidth 20 Mbps] { \label{cap_1:b} 
\includegraphics[width=5cm]{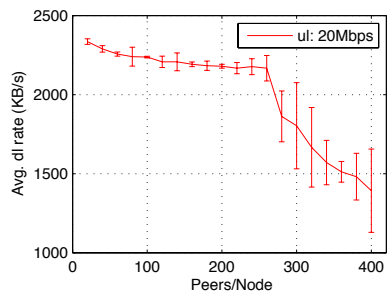} 
}

\subfigure[Upload bandwidth 40 Mbps] { \label{cap_1:c} 
\includegraphics[width=5cm]{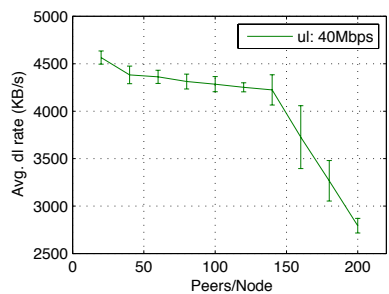} 
} 
\subfigure[Upload bandwidth 100 Mbps] { \label{cap_1:d} 
\includegraphics[width=5cm]{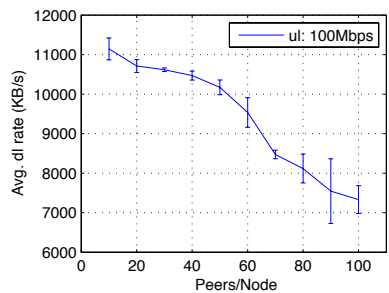} 
}
\caption{Average download rate as function of peers per node for
  different upload bandwidths for case of 1 node being used.}
\label{cap_1} 
\end{figure}

\begin{figure}[!tb]
\begin{center}
\includegraphics[width=8cm]{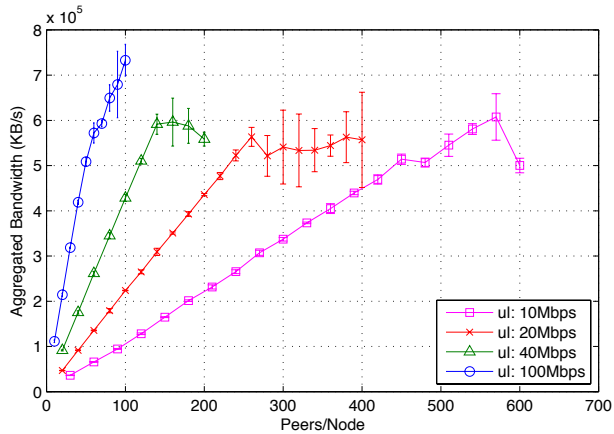}
\caption{The corresponding aggregated download bandwidth for the cases
  shown in Figure~\ref{cap_1}.}
\label{cap_2}
\end{center}
\end{figure}

Figure~\ref{cap_1} plots the average download rate against the number
of peers. Even though the maximum upload rates are set to different
values, the shapes of the curves are similar. The average download
rate decreases slightly until it reaches 40 peers per node, then it
enters into a relatively stable stage. After reaching the system
capacity, the average download rate drops sharply.

Figure~\ref{cap_2} plots the corresponding aggregated download
bandwidth based on the same experiment, with the curves from the
different cases combined. As the two figures show, the aggregated
download bandwidth can increase to 500~MB/s, and the corresponding
average download rates remain stable. Thus we can define 500~MB/s of
aggregated download bandwidth as the system capacity, and any value
below that is considered safe. Since the curves in the safe region are
basically straight lines, it is easy to fit a curve and we obtain the
result that relation between the number of peers per node, $x$, and maximum
upload rate per peer, $y$, is given by
\begin{equation}
  \label{eq:1}
  y = \frac{560}{x}.
\end{equation}

\begin{figure}[!tb]
%\newline
\begin{center}
\includegraphics[width=8cm]{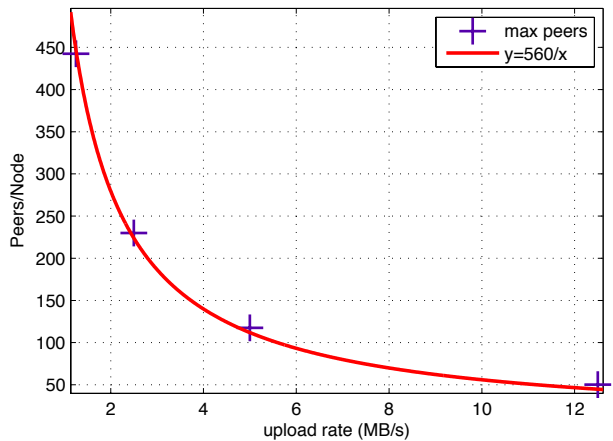}
\caption{Number of peers/node vs. per-peer upload rate.}
\label{cap_3}
\end{center}
\end{figure}

Figure~\ref{cap_3} shows this curve and our data points. This curve
can be used to set the values for upload bandwidth and number of peers
in an experiment \textit{when all peers are placed on a single
  node}. In download-constrained experiments, we get similar results
as in the upload-constrained experiments.

\subsection{Using More than One Node}
\label{sec:using-more-than}

\begin{figure}[!tb] 
  \centering 
  \subfigure[Average download rate] { \label{cap_4:a} 
    \includegraphics[width=5cm]{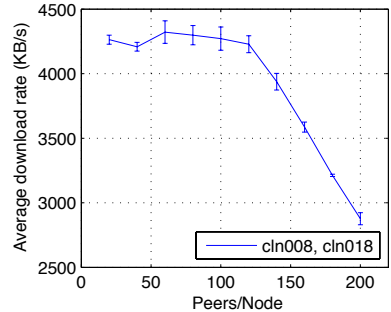} 
  } 
  \subfigure[Aggregated download bandwidth] { \label{cap_4:b} 
    \includegraphics[width=5cm]{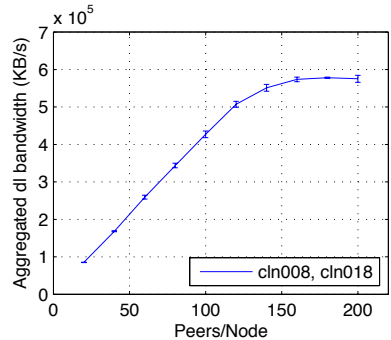} 
  } 
  \caption{Leechers deployed equally on two nodes; upload-constrained experiment}
  \label{cap_4} 
\end{figure}

We repeated the experiment above by placing the peers on two nodes
equally, starting with 20 peers per node and increasing by 20 peers
per node until we reached 200 peers per node. Upload rates were
constrained to 5~MB/s and download rates were unconstrained. Results
for average download rate and aggregated bandwidth are shown in
Figure~\ref{cap_4} and are at first sight similar to the ones obtained
for the single node case (Figure~\ref{cap_1:c} and 40~Mbps line in
Figure~\ref{cap_2}). 

In a download-constrained experiment, we obtained similar curves (not
shown due to space reasons).

Using the naive method, we would be led to conclude that 120 peers per
node is still safe, but when we inspected the actual network traffic
and connections made by the peers, we noticed that already at 60 peers
per node, the network between the nodes had been saturated (see
details below). As a result of this saturation, BitTorrent changed its
behavior. This is not apparent in Figure~\ref{cap_4}, hence the naive
method is inadequate. Below we provide details about our observations
of the changes in behavior and an analytical means for calculating
when experiments are still safe.

However, we consider the lack of observed change in the average
download rate in changing network conditions as excellent evidence of
BitTorrent's ability to adapt to varying conditions. 

% As the figure \ref{cap_4} shows, when we deploy the leechers on 2 nodes, the whole system still exhibit a capacity of more than 500MB/s. And before 120 peers/node, the average download rate remains stable. However, by monitoring the network traffic, we found the network has already been saturated when the peers/node reaches 60. Figure \ref{cap_4} cannot exhibit the change in BitTorrent behaviors. That's the reason why we call this method naive and average download rate cannot be used as the only benchmark when we design the experiment (especially when more nodes are used for deploying leechers).

% In the following section, we will introduce a better way to calculate capacity. The behavior change will be discussed in details in section \ref{cluster_nodes}.

\subsection{Clustering of the Nodes}
\label{cluster_nodes}

We will substantiate our above claim that BitTorrent's behavior has
changed and that the average download is not an accurate indicator of
a correct experiment in two ways. First, we will experimentally
investigate how the connections between the peers are formed in the
above experiment. Second, we derive simple analytical expressions for
determining the bounds of when an experiment can be considered
correct, and demonstrate that the above experiment with two nodes
violates these intuitive conditions.

% According to our formulas for capacity planning, the system should
% reach its limit at the point 60(??). It means when there are 60(??)
% peers on each of the two nodes, the network bandwidth will be
% saturated. However, the average download rate still remains at a
% stable level at that point. It starts to drop at the point 160, before
% that it remains stable and the aggregated bandwidth keeps increasing
% steadily. That is why we claim average download rate and aggregated
% bandwidth cannot be used as the only benchmark for capacity planning,
% since it cannot reflect the real system capacity and hide the changes
% of BitTorrent's behaviours.

% We claim the BitTorrent has changed it behaviours at the point 60(??),
% which cannot be reflected by the average download rate. The change of
% the behaviours originates from peer selection strategy. Since the
% network bandwidth has been saturated, a peer can have better download
% rate from native buddies. According to peer selection strategy, a peer
% will show preference in exchanging data with native buddies. In other
% words, they tend to cluster.

We ran the experiment with two nodes as above, i.e., start with 20
peers per node, increasing it by 20 peers per node until we reach 200
peers per node. Upload rates were constrained to 5~MB/s and download
rates were unlimited. In every experiment we kept track of all the
connections maintained by all the peers and identified which
connections are \textit{native} (to peers on same node) and which are
\textit{foreign} (to peers on the other node). Every experiment was
repeated 3 times and we present the averages and the standard
deviations in the figures.

% In order to testify our claim and formulas, we designed the
% following experiments. In the first round experiments, we deployed
% the leechers on two nodes equally. We started from 20 peers/node,
% and increased 20 peers on each node in every following experiment
% until it reached 200 peers/node. The max upload rate of every peer
% is set to 5000KB/s, and no limits on download rate. In every
% experiment, we kept track of all the connections maintained by a
% peer. Then we calculated how many upload connections are to the
% peers on the same node among all the uploads connections. Every
% experiment are repeated 3 times and the average value is used. The
% statistics are obtained from the snapshot function in our Logger
% module.

\begin{figure}[!tb]
  \begin{center}
    \includegraphics[width=8cm]{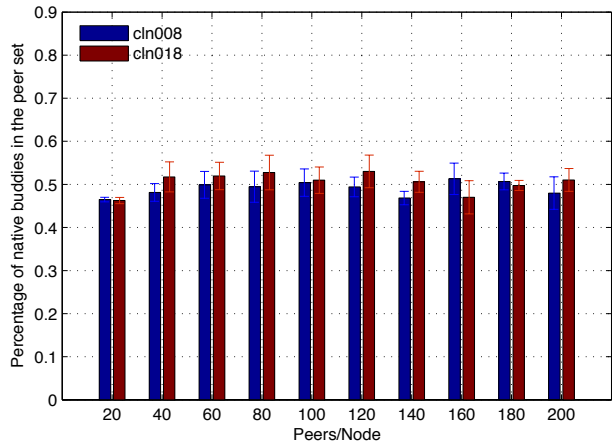}
    \caption{Fraction of native buddies in the peer list}
    \label{connr_1}
  \end{center}
\end{figure}

Figure~\ref{connr_1} shows the fraction of native buddies in the peer
list given by the tracker. As we can see, the value hovers around 50\%
which is to be expected since the tracker picks the peers for the peer
list uniformly at random. Investigating the fraction of native buddies
(and consequently foreign buddies) allows us to determine how
BitTorrent is choosing where to download from.

% The figure \ref{connr_1} shows the composition of a peer's buddies,
% the percentage of the buddies on the same node with the peer among
% all the buddies it maintained. We can see this value is always
% around 50\%. This is easy to explain. Since we deployed leechers on
% two nodes equally. And the tracker will make a peer-list by randomly
% selecting. There should be 50\% of the peers from the same node in a
% peer-list. That's why this value remains stable in all the
% experiments, it reflects the results from the peer-list.

\begin{figure}[!tb]
\begin{center}
\includegraphics[width=8cm]{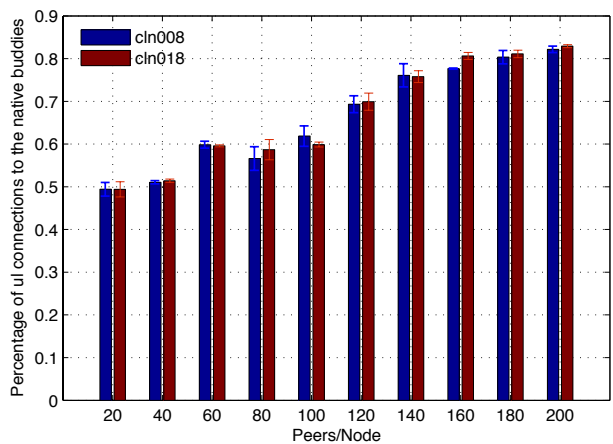}
\caption{Fraction of upload connections to native buddies in an
  upload-constrained experiment with 2 nodes}
\label{connr_2}
\end{center}
\end{figure}

Figure~\ref{connr_2} shows the fraction of upload connections to
native buddies in an upload-constrained experiment. We used two nodes,
\texttt{cln008} and \texttt{cln018} and show the values for both of
them, as a function of the number of peers per node. As we can see,
from 60 peers per node onwards, the peers tend to \textit{favor native
buddies} and the fraction of connections to native buddies keeps on
increasing throughout the experiment.

The explanation is quite simple. Because the peers obtained in the
peer list are evenly distributed, so are the connections in the
smaller tests. Because both the native and foreign peers are able to
serve data equally fast, a peer has no reason to prefer one over the
other. (Recall that BitTorrent selects the peers to upload to or
download from based on the bandwidth it obtains to/from that peer.) At
around 60 peers per node, the amount of data going between the nodes
is enough to saturate the 1~Gbps network link, whereas the local
loopback still has a lot of unused capacity. Hence, what we are seeing
in Figure~\ref{connr_2} is simply the normal BitTorrent's peer
selection algorithm at work. In other words, the peers have clustered
themselves locally but this effect is not visible in the average
download rates or aggregate bandwidth shown in Figure~\ref{cap_4}. 

% The figure \ref{connr_2} shows among all the upload connections, the
% percentage of the connections to the native buddies. We can see this
% figure confirms our prediction. Before 60(??) peers/node, the
% network is not saturated and the experiments are performed within
% the system capacity. A peer can obtain equally good download rate
% from native buddies and foreign buddies. A peer shows no preference
% in selecting buddies to upload. So half of the upload connections
% are made to the native buddies based on the probability.

% As the value of peers/node increases, more aggregated traffic will
% go through the ethernet card into the network, and the network will
% be saturated first. After 60(??) peers/node, a peer can obtain
% better download rate from native buddies, and will more likely
% upload to them as a result of tit-for-tat. The effects of peer
% selection strategy become more and more visible. When peers/node
% reaches 160, the aggregated traffic still not hit its peak value,
% but almost 80\% of the upload connections are made to peers on the
% same node already. This experiment confirmed our prediction well.

\subsection{Clustering with Download Constraints}
\label{cluster_nodes_dl}

We repeated the above experiment, but this time constrained the
download rate of every leecher to 5~MB/s and left the upload rates
unlimited. Seed's maximum upload rate was 5~MB/s as in the other
experiments. 

% Then we designed download-constrained experiments as our second
% round of experiments to reconfirm our prediction. The configuration
% of the experiments are basically the same as the first round, except
% the leechers' max download rate is set to 5000KB/s, and no limits on
% upload rate. The seed's max upload rate is still 5000KB/s.

\begin{figure}[!tb]
\begin{center}
\includegraphics[width=8cm]{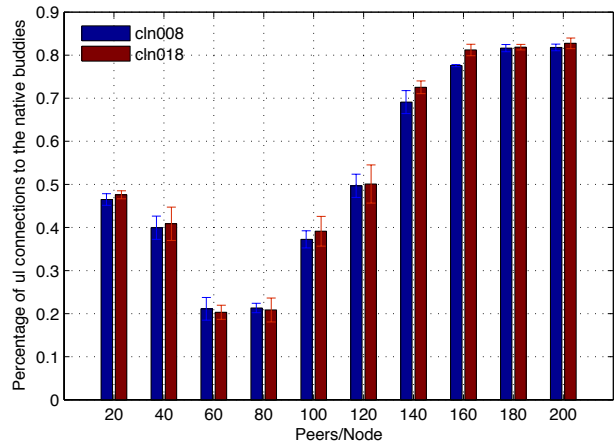}
\caption{Fraction of upload connections to native buddies in a
  download-constrained experiment with 2 nodes.}
\label{connr_3}
\end{center}
\end{figure}

Figure~\ref{connr_3} shows the results from this experiment. As with
the upload-constrained case, the network is saturated at around 60
peers per node, but the effects are drastically different from the
upload-constrained case. The peers start favoring \textit{foreign
  buddies} as opposed to native buddies for a longer spell and return
to favoring native buddies only in very large experiments.

% The figure \ref{connr_3} shows the experiment result. We can see the
% figures are quite different from those in the first round
% experiments. The most significant difference is at the point
% 60(??). Our formula shows the network is already saturated at that
% point, so BitTorrent changes its behaviours accordingly. The peers
% tend to select the native buddies to upload in the
% upload-constrained experiments. However, in the download-constrained
% experiments, they tend to upload to foreign, why?

% By tracking the amount of data received by node cln018, we found
% even 80\% upload connections are made to the foreign buddies, most
% of the data received by cln018 are contributed by native buddies,
% which can be seen in table XXX. This gave us a very good clue to
% analyse the experiment results.

% \begin{table}[!tb]
% %\renewcommand{\arraystretch}{1.0}
% \centering
% \begin{tabular}{|l|l|l|l|l|l|l|l|l|l|l|}
% \hline
% $cln008$ & 20 & 40 & 60 & 80 & 100 \\
% \hline
% Native& 54.4\% & 56.9\% & 60.1\% & 68.4\% & 77.2\% \\
% \hline
% Foreign& 45.6\% & 43.1\% & 39.9\% & 31.6\% & 22.8\% \\
% \hline
% \end{tabular}
% \caption{Percentage of traffic through the loopback and percentage of traffic through eth0 as a function of different peers/node, corresponding to the figure \ref{connr_3} }
% \label{tab:table_cln008_traffic}
% \end{table}

Interestingly, our analysis of the situation showed that although most
of the upload connections in the range of 60--80 peers per node were
to foreign buddies, the peers received most of the data from native
buddies. For example, with 20 peers per node, 54.4\% of the traffic
came from native buddies, at 60 peers per node this was 60.1\% and at
100 peers per node 77.2\%. Turns out that this behavior is a result of
BitTorrent's \textit{piece selection strategy}. Piece selection
strategy in BitTorrent is based on a mechanism called
rarest-first. The purpose is to make a peer attractive to the others
by requesting the rarest pieces first in the swarm, and quickly turn a
peer into a productive member of the swarm.

% Our analysis shows the significant change in BitTorrent's behaviours at the point 60(??) is the result of its piece selection strategy. Before we go into the analysis, we should take a look at BitTorrent's piece selection strategy. Piece selection strategy in BitTorrent is mainly based on a mechanism called rarest-first. The purpose is to make a peer attractive to the others by requesting rarest pieces first in the swarm, and quickly turn a peer into a productive player. 

Peers make the decision on which piece they consider to be the rarest
based on locally available information from other peers. (This is why
in some cases BitTorrent's piece selection algorithm is called local
rarest first.) Peers obtain information about the pieces other peers
possess through BitTorrent's HAVE-control messages. A peer sends a
HAVE-message to its buddies when it has completed the download of a
piece, in order to let its buddies know that they can download the
piece from the peer. Peers keep track of the HAVE-messages and use
them to calculate which pieces are the rarest among their buddies.

% Then how can a peer decide a piece is rare? It is impossible for a peer to have the global information of the piece distribution in the swarm, so the decision for the rarest piece must be decided based on the local information. That’s why some people also refer it as Local Rarest-First. The solution is the HAVE message. We know that a peer will use HAVE message to inform its buddies whenever it receives a complete piece. In such a way, a peer can keep track of the number of HAVE messages for each piece. The piece with the least HAVE messages from its buddies is the rarest.

BitTorrent's control messages (of which HAVE is one) have to share the
network with the actual data transfers. When the network (or loopback
device) becomes congested, both the data and control messages are
slowed down. At the 60 peers per node point, the network between the
nodes starts becoming congested, but the loopback is still far below
its capacity. Hence, peers receive a lot of HAVE-messages from the
native buddies but the HAVE-messages from foreign buddies slow
down. As a result of this, the peer (correctly) considers the pieces
from the foreign buddies to be rarer than native pieces (which spread
very fast within the node to many peers) and wants to request the
rarer pieces from the foreign buddies first. As the network is only
approaching the saturation point and is not yet completely congested,
the peer is able to provide uploads to foreign buddies so that they
are willing to upload pieces to it (recall the use of tit-for-tat).

% The messages exchanged among the peers are divided into two categories, one is control messages(HAVE message is one of them), the other data messages. What's more, the messages exchanged between the peers on different nodes will go through ethernet card; while the messages exchanged between the peers on the same node will go through loopback device. When the network becomes congested, not only the data messages will be slowed down, but also the control messages. However, at the point 60(??), the aggregated traffic through loopback is far below its limit, both control and data messages spread very fast within the node. This causes a peer receive more HAVE messages for each piece from the native buddies than foreign buddies. As a result, a peer will think the pieces possessed by foreign buddies are the rarest, and will request them first. Even though at this point, a peer cannot provide its buddies on the other node as high upload rate as before, but not bad enough to be choked yet. It still can be served.

From the results, we can conclude that BitTorrent's piece selection
algorithm is very sensitive to changes in network conditions in the
download-constrained cases. In fact, piece selection strategy
overrules peer selection strategy in the early part of the
experiment. As the network gets more and more congested, the peers are
no longer able to provide good enough upload rates to foreign buddies,
so in accordance to the tit-for-tat policy, they are choked. Hence
they have to resort to the native buddies for actually getting the
data. Since there are no limits on upload rate, the actual injection
of new information is limited by the seed's upload rate (which was
limited), but the native buddies are enough to feed new data within
the node. Eventually, we see the same kind of clustering between peers
on a single node as we saw in the upload-constrained case.

\begin{figure}[!tb]
\begin{center}
\includegraphics[width=8cm]{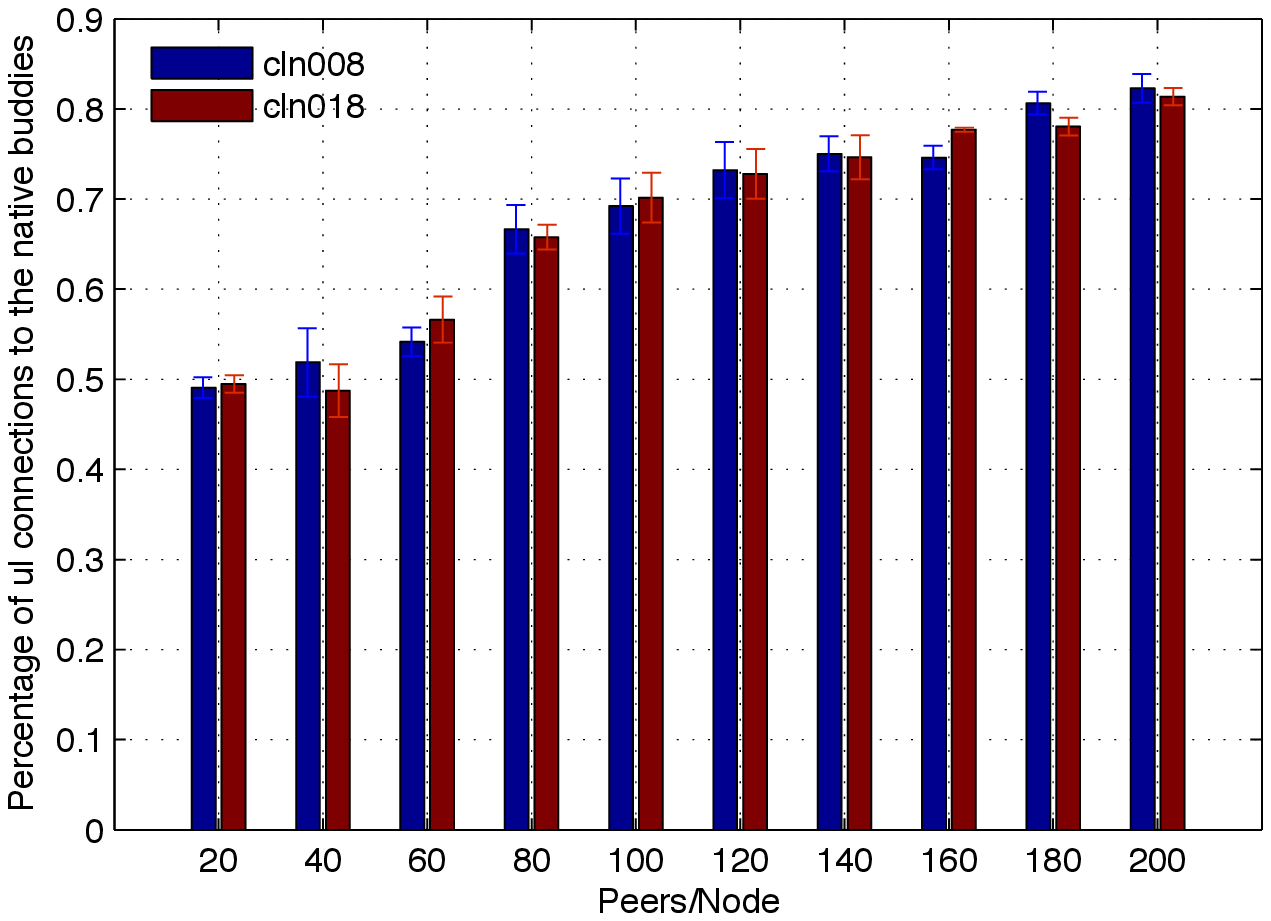}
\caption{Fraction of upload connections to native buddies in a
  download-constrained experiment with 2 nodes and random piece selection.}
\label{connr_4}
\end{center}
\end{figure}

To verify our claim that the behavior above is due to the piece
selection algorithm, we repeated the experiment with a random piece
selection algorithm. Because peers exhibit no preference for pieces,
peer selection algorithm should be the deciding factor. Results are
shown in Figure~\ref{connr_4}. The results are similar to the
upload-constrained case in Figure~\ref{connr_2} where peer selection
is known to be the deciding factor.

% In order to verify our analysis for the download-constrained experiment, we designed the third round of experiments. In the third round experiments, we substitute random-selection for rarest-first as our piece selection strategy. The other experiment settings are exactly the same as those in the second round.

% Our previous analysis claims piece selection has greater influence on BitTorrent's behaviours in the beginning of the network congestion, and further causes the differences between download- and upload-constrained experiments. If our analysis is right, by replacing rarest-first with random-selection, we should observe the significant influence from rarest-first vanishes, only the the influence form peer selection will be visible in the figure. Since random-selection shows no preference in selecting which piece to request.

% From the figure \ref{cap_4}, we can see the experiment results confirmed the correctness of our analysis. The BitTorrent shows the same behaviours at those in upload-constrained experiments.

% Actually, we also did another two rounds experiments, in one of which, both leechers' max upload and download rate are set to 5000KB/s, then we got the same figure as the one in the first round experiment; in the other, both max upload and download rate are set to unlimited, then we got the same figure as the one in the second round experiment. Anyway, all the previous experiments verified our formulas and prediction from different angels.

% At the point where the network capacity is just saturated, piece selection has very strong influences on BitTorrent behaviours. It is so strong that even in upload-constrained experiment. We can still observe the its influence at the point 60 MB/s. However the peer selection strategy dominates the BitTorrent's behaviour in upload-constrained experiment. Especially in the latter part of the figure.

% Another strong argument is if we don't limit the concurrent uploads to 7, but set it to 80, which means all the buddies will be served at the same time. We repeated the upload-constrained experiment. As the figure XXX shows, 

\begin{figure}[!tb]
  \begin{center}
    \includegraphics[width=8cm]{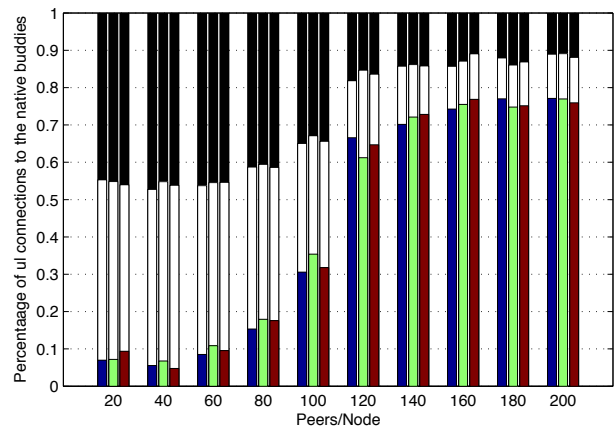}
    \caption{Fraction of upload connections to native buddies (lowest
      section of bars) and foreign buddies (white and black sections)
      with 3 nodes in a download-constrained experiment.}
    \label{connr_5}
  \end{center}
\end{figure}

As further evidence, we ran the download-constrained experiment with
leechers placed equally on three nodes and the fraction of connections
to other nodes is shown in Figure~\ref{connr_5}. The three parallel
bars represent the three nodes. The lowest section of each bar shows
native connections and the two upper sections show connections to the
two other nodes. We see the same preference for foreign buddies in the
beginning, with connections between the other two nodes being rather
uniformly split, as is to be expected. After the network gets
congested, we see the same kind of clustering as in the case with two
nodes. 

\subsection{Capacity Planning Formulas}
\label{sec:capac-plann-form}

We now present a simple analytical means of determining whether a
planned experiment falls within the system capacity limits or not.
Table~\ref{tab:table_cpvar} lists the notation used in the
following. Let $i, j, k\in\lbrace1,2,3...n \rbrace$. Compared with the
traffic among the leechers, the traffic from the seed is negligible
and we have excluded it for reasons of simplicity.

\begin{table}[!tb]
\renewcommand{\arraystretch}{1.0}
\centering
\begin{tabular}{|l|p{5.5cm}|}
\hline
$n$ &  number of nodes in an experiment\\
\hline
$m_i$ & number of peers on node $i$\\
\hline
%$u_{ij}$ & upload bandwidth of ????\\
%\hline
%$d_{ij}$ & download bandwidth of ????\\
%\hline
$U_i$ &  aggregated upload bandwidth generated by the peers on node $i$\\
\hline
$D_i$ &  aggregated download bandwidth generated by the peers on node $i$\\
\hline
$L_i$ & physical capacity of loopback device on node $i$\\
\hline
$C^{ul}_i$ & physical upload capacity of network card on node $i$\\
\hline
$C^{dl}_i$ & physical download capacity of network card on node $i$\\
\hline
%$P_{i}$ &  probability that a peer on node $i$ will connect to native peers \\
%\hline
%$P_{i}^{\prime}$ & probability that a peer on node $i$ will connect to foreign peers \\
%\hline
$P_{ij}$ &  probability that a peer on node $i$ will connect to peers on node $j$ \\
\hline
\end{tabular}
\caption{variables used in the discussion}
\label{tab:table_cpvar}
\end{table}

When we deploy multiple peers on one node, a peer will not only try to
connect and upload to the native peers, but also to the foreign
peers. $P_{ij}$ is the probability that a peer on node $i$ will
connect to peers on node $j$, and assume all the peers on node $i$
have the same probability. Then we have

\begin{equation}
P_{ij} = \left\{
\begin{array}
{rl}
\frac{m_{i} - 1}{\sum_{k=1}^{n} m_k - 1 } & \text{if } i = j ,\\\\
\frac{m_j}{\sum_{k=1}^{n} m_k - 1} & \text{if } i \neq j .
\end{array}
\right. 
\end{equation}

When $i=j$, $P_{ii}$ actually denotes the probability that a peer will
connect to the native peers. 

%\texttt{Let $P_{i}^{\prime}$ denote the probability that a peer on node $i$ will connect to any foreign peers, and assume all the peers on node $i$ have the same probability.  Then we have}

%\begin{equation}
%P_{i}^{\prime}  = 1 - P_{i} = \frac{ \sum_{k=1, k \neq i}^{n} m_k }{\sum_{k=1}^{n} m_k - 1} \label{eq_pother}
%\end{equation}

$U_i$ and $D_i$ denote the aggregated upload and download bandwidth on
node $i$ respectively. Obviously, $U_i$ equals the sum of all peers'
upload bandwidth on node $i$ and $D_i$ equals the sum of all peers'
download bandwidth on node $i$. Then the traffic from node $i$ to node
$j$ is\footnote{We have made the assumption that all peers on a node
  have the same limits on upload and download bandwidths.}

\begin{equation}
T_{ij} = P_{ij} \times min(U_i, D_j)
\end{equation}

We can construct a matrix to show the traffic flows between the nodes:
\begin{equation}
T = 
\begin{bmatrix}
T_{11} & T_{12} & T_{13} &\ldots &  T_{1n} \\
T_{21} & T_{22} & T_{23} & \ldots  & T_{2n} \\
\vdots & \vdots & \vdots & \ddots & \vdots \\
T_{n1} & T_{n2} & T_{n3} & \ldots  & T_{nn} \\
\end{bmatrix}
\end{equation}

In matrix $T$, row $i$ represents the distribution of the traffic
flowing out of node $i$, and column $i$ represents the distribution of
the traffic flowing into node $i$. The elements on the diagonal
represent the traffic going through the loopback interface of a
node. This traffic in $T$ must be constrained by the physical capacity
of a node. Then for each $i, j$ %$i, j \in\lbrace1,2,3...n \rbrace$,
we have:

\begin{equation}
\sum_{i=1, i \neq j}^{i=n} T_{ij} \leqslant  C_{j}^{dl} \label{eq_ucap1}
\end{equation}

\begin{equation}
\sum_{j=1, i \neq j}^{j=n} T_{ij} \leqslant  C_{i}^{ul} \label{eq_ucap2}
\end{equation}

\begin{equation}
T_{ii} \leqslant  \frac{L_{i}}{2} \label{eq_ucap3}
\end{equation}

Now, consider an extreme situation, when all the traffic goes through
loopback interface or the network card, then we have the following
constraints: 

\begin{equation}
U_{i} \leqslant  C_{i}^{ul} \label{eq_lcap1}
\end{equation}

\begin{equation}
D_{i} \leqslant  C_{i}^{dl} \label{eq_lcap2}
\end{equation}

\begin{equation}
min(U_{i}, D_{i}) \leqslant  \frac{L_{i}}{2} \label{eq_lcap3}
\end{equation}

To some extent, \eqref{eq_ucap1}, \eqref{eq_ucap2} and
\eqref{eq_ucap3} define the upper bound of the experiment, while the
\eqref{eq_lcap1}, \eqref{eq_lcap2} and \eqref{eq_lcap3} define the
lower bound. The upper and lower bound will converge at two
points. The first is when only one node is used for deploying
leechers. Then there is only one element $T_{11}$ in the matrix. The
\eqref{eq_ucap3} and \eqref{eq_lcap3} will be the same, since all the
traffic will go through the loopback interface.

The second is when an infinite number of nodes is used. Considering
that we can only deploy a limited number of peers on a node, the
probability that a peer will connect to native peers decreases to
zero. As a result, all the traffic will go through network card. Then
\eqref{eq_ucap1} and \eqref{eq_ucap2} will be the same as
\eqref{eq_lcap1} and \eqref{eq_lcap2}. $T_{ii}$ will be zero since no
traffic will go through the loopback interface.

% There are several other things worth remarking here. Our analysis above are under the assumption that all the peers are started up at the same time, and only applies to the beginning of an experiment. Since at that time, peer selection strategy has not taken effects yet, the traffic will be distributed merely on the probability we have calculated above. However, at time goes by, the traffic may shift because of the experiment settings and the BitTorrent's internal mechanisms. How and how fast will the traffic shift occur depends on the specific experiment configurations.

% One thing for sure is as more nodes are involved in the experiment, the capacity planning will become easier. Basically,  \eqref{eq_lcap1} and \eqref{eq_lcap2} will decide the experiment scale.

\subsection{Example: Case of 2 Nodes }
\label{sec:example:-case-2}

We revisit the case of using two nodes in an experiment shown in
Figure~\ref{cap_4}. In the experiment, we obtained an average download
rate of 4.25~MB/s and loopback capacity $L_i = 500 $~MB/s. The network
between the nodes is a Gigabit Ethernet, so $C_{i}^{ul}=125MB/s$ and
$C_{i}^{dl}=125MB/s$. ($i \in \lbrace 1, 2\rbrace$)

When there are 40 peers on each of the two nodes, we get the traffic
distribution matrix $T^{40}$ as below:

\begin{equation}
\label{eq:pn40}
T^{40} = 
\begin{bmatrix}
T_{11} & T_{12} \\
T_{21} & T_{22} \\
\end{bmatrix}
=
\begin{bmatrix}
83.9 & 86.1 \\
86.1 & 83.9 \\
\end{bmatrix}
\end{equation}

We can see from the equation \eqref{eq:pn40}, for node 1,
$T_{12}~\leqslant~C_{1}^{ul}$, $T_{21}~\leqslant~C_{1}^{dl}$ and
$T_{11}~\leqslant~\frac{L_{1}}{2}$. The same applies to node 2. We can
see all the equations hold, the experiments are designed within the
system capacity.

%What's more, since it is an upload-constrained experiment, we should not observe any clustering phenomenon.

When there are 60 peers on each of the two nodes, we obtain the traffic
distribution matrix $T^{60}$ as below:

\begin{equation}
\label{eq:pn60}
T^{60} = 
\begin{bmatrix}
T_{11} & T_{12} \\
T_{21} & T_{22} \\
\end{bmatrix}
=
\begin{bmatrix}
126.4 & 128.6 \\
128.6 & 126.4 \\
\end{bmatrix}
\end{equation}

We can see from the equation \eqref{eq:pn60}, for node 1,
$T_{12}~\textgreater~C_{1}^{ul}$ and
$T_{21}~\textgreater~C_{1}^{dl}$. Both equations \eqref{eq_ucap1} and
\eqref{eq_ucap2} are violated. Since equation \eqref{eq_lcap3} still
holds, then in a upload-constrained experiment, a peer will not treat
native buddies and foreign buddies equally. They start showing
preference in uploading to native buddies, and the clustering
happens. The same analysis can be applied to node 2. This analysis
yields the same result as the investigation on the actual behavior of
BitTorrent above.

%In a nutshell, the experiment is not properly designed if 60 peers are deployed on each of the two nodes. Since the clustering phenomenon is NOT because the  

%The calculation matches the results from the experiments in section \ref{cluster_nodes} very well.

% \section{Discussion}
% \label{sec:discussion}
% The difference between cluster and internet.
% However, as far as cluster is concerned. 32KB is still not enough. Besides the short RTT and low packet loss rate, high bandwidth delay product is also one of the characteristics of a cluster. Usually the TCP is tuned in a cluster to guarantee the performance. 

% Other general discussion about the applicability of clusters in
% experiments.

\section{Related Work}
\label{related_work}

BitTorrent has been a popular target for research over the past
several years, including several papers, e.g.,~\cite{meu09, ross06,
  sirvian07, legout07, legout05} that use a real BitTorrent client in
their experiments to validate their models and conclusions. However,
less papers have concerned themselves with the accuracy of their
experiments and possible bias in their methodologies.

Legout et al.~\cite{legout05, legout06} have made a thorough
measurement-based research on the two core mechanisms of BitTorrent,
piece and peer selection.. However, the influences from these two
mechanisms are discussed separately. The authors showed that the
rarest first algorithm guarantees a close to ideal entropy, while the
choke algorithm guarantees the fairness in the system. None of the
results presented in the papers investigate the combined effects of
the two mechanisms, which as we have shown, also occurs and can have
significant effects on BitTorrent's behavior.

Antoniu et al.~\cite{antoniu04} discuss the difficulties in validating
large-scale peer-to-peer systems. The authors also proposed a
framework for performing large-scale experiments based on grid
services. However, how the experiments are affected by the underlying
details and the experiment settings are not touched.

However, only a few papers, e.g.,~\cite{ras07,boxun10,rao10} concern
the accuracy of experiments and the bias of measurements. Work
in~\cite{boxun10} investigated the sampling bias in BitTorrent
experiments. Even though the discussion merely focuses on the approach
of using instrumented client to obtain data from real-world swarm, the
recommendations proposed in this paper are simple heuristics and
guidelines. We have followed their recommendations and have designed
our Logger module to follow them. Our Logger module takes a snapshot
for the peer every second during its whole life span. This strategy
yields very reliable experiment data.

On the other hand, Rasti and Rejaie~\cite{ras07} claim that the data
obtained with this approach (injecting an instrumented client into
real-world swarm) is not representative and has already been biased in
the beginning. The main reason for their claim is that BitTorrent
clients tend to cluster with other clients having similar upload
bandwidths. This observation is definitely valid for measuring a
real-world swarm on the Internet, but as our experiments are performed
on a cluster where all peers are instrumented to provide logging
information, such a bias does not exist in our experimental setup.

A lot of analytical work has also studied the clustering properties of
BitTorrent. Based on the analysis of the choking
algorithm,~\cite{legout07} provides empirical evidence of BitTorrent's
clustering and show that peers with similar bandwidths tend to get
clustered. 

Meulpolder et al.~\cite{meu09} extend an earlier analytical model
from~\cite{qiu04} and propose a new model for analytical investigation
of BitTorrent's clustering. Their model only takes into account peer
selection in BitTorrent and ignores the effects of piece
selection. They observe similar clustering behavior as we have
observed. However, their model and measurements exhibit a small
discrepancy which they conjecture is the result of probabilistic
effects from too small experiments. Our results show that clustering
in BitTorrent is actually an interplay of both peer and piece
selection algorithms, and we believe that their observed discrepancies
are a result of their model ignoring piece selection. Although the
effects of piece selection on clustering are small and hard to
observe, our work, in particular on the download-constrained
experiments, has shown that it cannot be ignored. Both~\cite{meu09}
and our work find the same effect of upload connections going to
foreign peers while the majority of data comes from native peers.

% Based on the intensive analysis on choke algorithm, \cite{legout07} gives the empirical evidence of BitTorrent's clustering property. The author claimed peers with similar bandwidth are easy to get clustered. \cite{meu09} extends the analytical model in \cite{qiu04}, proposes a new model to analyse the clustering property. Even though BitTorrent exhibited quite different clustering property caused by piece selection in some of our experiments. The unlimited upload-bandwidth is not realistic in common sense. Our research shows the influences from piece selection can be neglected when modelling clustering property.

The work by Rao et al.~\cite{rao10} is the closest work to ours. The
authors discuss the rationality of performing BitTorrent experiments
on a cluster. However, the discussions focus on the marginal
influences on the average download rate from various RTT and packet
loss rates and conclude that the effects from changing RTTs and packet
loss rates are so small that they can be discounted in the
evaluation. Our work focuses on how to design an experiment on a
cluster properly, i.e., what is the ``safe region'' for a correct
experiment and how BitTorrent behaves when the experiments are
performed around the system capacity limit.

The experiment setup in~\cite{rao10} is very similar to the case
discussed in our paper. The authors used 3 nodes for deploying
leechers (100 leechers on each node) and performed a homogeneous
upload-constrained experiment. The maximum upload rate was set to
100~KB/s. They did not consider possible bottlenecks in their
experiment setup. Using our capacity planning method from
Section~\ref{sec:capac-plann-form}, we can see that their experiments
require only on the order of 3~MB/s of bandwidth between nodes and on
the loopback. Given that they were using modern computers on the Grid
5000 testbed, they should be well below the system capacity limit. Our
work therefore validates their experiment setting as being correct.

% We applied our capacity planning method and construct the
% traffic distribution matrix for it as below (unit: KB/s):

% \begin{equation}
% \label{eq:rao200}
% T^{200} = 
% \begin{bmatrix}
% T_{11} & T_{12} & T_{13} \\
% T_{21} & T_{22} & T_{23} \\
% T_{31} & T_{32} & T_{33} \\
% \end{bmatrix}
% =
% \begin{bmatrix}
% 3311 & 3344 & 3344 \\
% 3344 & 3311 & 3344 \\
% 3344 & 3344 & 3311 \\
% \end{bmatrix}
% \end{equation}

% Even the author of \cite{rao10} didn't specify the capacity of loopback interface and network bandwidth explicitly. The elements in $T^{200}$ are quite small. So all of our capacity planning formulas should hold in normal situations. The experiment design is reasonable.

\section{Conclusion}
\label{conclusion}

Experimental evaluation of large scale systems is an important topic
in networking research. Currently no ideal environment exists for such
evaluations, with simulations, real Internet, and cluster-based
testbeds being the commonly used solutions. We believe that
cluster-based testbeds offer the best of both worlds, realistic
applications with a real (albeit not necessarily realistic) network in
between. 

In this paper we have shown how to design BitTorrent experiments on a
cluster. Our focus has been on identifying how the physical limits of
the host machine affect the tests and how many clients can be deployed
on a node. We have shown that the number of peers per node depends on
many factors, but up to 500~peers per node is realistic for certain
values of allocated per-client bandwidth. We have shown that the
simple metric of average download rate is not sufficient for
determining when an experiment is ``safe'', but that a more complex
analysis is needed. We provide a simple set of formulas, intended to
be used as rules of thumb for determining if an experiment runs into
the physical limits of the machine. 

Our work has also extended previous work on BitTorrent, by showing
that the previously observed clustering behavior is actually a result
of both the peer and piece selection algorithms, and not simply the
peer selection algorithm as previously believed. Although the effect
of the piece selection algorithm is small, it cannot be ignored in all
cases. 

In our future work, we plan to verify our results using a 10~Gbps
network between the nodes. This is likely to change some of the
details of our results, since in that case the loopback will saturate
before the network; hence the clustering behavior will be different. 

% Recommendations: 

% An experiment should be performed in the "safe region". The closer an experiment approaches to the system capacity, the more an experimenter will be haunted by various parameters. Keep the experiment design in the safe region can free the researchers from the underlying details and focus on the problem itself.

% When running multiple peers on one node, use as many nodes as possible in the experiment. Then the influence from loopback device can be neglected. When running experiment in 10Gbit ethernet, capacity planning is easy.

% Low transmission rate and large distribution file are always safe choices. Be very careful when using high transmission rate, high transmission rate and small distribution file are bad choices.

% In homogeneous experiment, limiting the upload bandwidth is more important than limiting the download bandwidth. In heterogeneous experiment, limiting the download rate is necessary.

\bibliographystyle{IEEEtran}
\bibliography{lahteet}

% that's all folks
\end{document}